\documentclass[twocolumn,a4paper,showpacs,preprintnumbers,amsmath,amssymb,floatfix]{revtex4-1}

\pdfoutput=1
\usepackage{graphicx}
\usepackage{dcolumn}
\usepackage{bm}
\usepackage{times}
\usepackage{multirow}
\usepackage{longtable}
\usepackage{xcolor}

\begin{document}


\title{Spatially resolved Landau level spectroscopy of the topological Dirac cone of bulk-type Sb$_2$Te$_3$(0001): potential fluctuations and quasiparticle lifetime}
\author{C. Pauly$^1$}
\email[] {pauly@physik.rwth-aachen.de}
\author{C. Saunus$^1$}
\author{M. Liebmann$^1$}
\author{M. Morgenstern$^1$}
\affiliation{ $^1$II. Physikalisches Institut B and JARA-FIT, RWTH Aachen University, D-52074 Aachen, Germany}

\date{\today}

\begin{abstract}
Using low temperature scanning tunneling spectroscopy, we probe the Landau levels of the topologically protected state of Sb$_2$Te$_3$(0001) after in-situ cleavage of a single crystal. Landau levels are visible for magnetic fields $B\ge 2$\,T at energies, which confirm the Dirac type dispersion including the zeroth Landau level. We find different Dirac velocities for the lower and the upper part of the Dirac cone in reasonable agreement with previous density functional theory data. The Dirac point deduced from the zeroth Landau level shifts by about 40\,meV between different areas of the sample indicating long range potential fluctuations. The local potentials are correlated to different local defect densities varying slightly stronger than expected from a statistical distribution. Moreover, the width of the Landau level peaks is analyzed. It is found to increase, mostly linearly, with the energy distance to the Fermi level. Consequently, we attribute the peak width to a dominating scattering of the hot quasiparticles by electron-electron interaction.
\end{abstract}

\maketitle

\section{Introduction}
Topological indices are a new paradigm to classify solids by relating bulk properties unequivocally to the conductivity at the rim of the sample~\cite{Kane,Schnyder,Hasan,Qi,Yan}. They have been used to categorize the quantum Hall effect~\cite{Thouless,Kohmoto} as well as to predict a non-magnetic quantized transversal conductance in superfluids \cite{Volovik}. More recently, they led to the experimental discovery of two-dimensional (2D) topological insulators (TIs)~\cite{Konig,Volkov,Kane2}, strong and weak three-dimensional (3D) TIs~\cite{Fu,Hsieh,Zhang,Ringel,Rasche,Pauly}, topological crystalline insulators~\cite{Fu2,Dziawa} and the anomalous quantum Hall effect~\cite{Yu,Chang}. The strong 3DTIs are so far the most versatile class in terms of different realizations in materials~\cite{Hasan,Qi,Yan,Zhang2,Yang}. The driving force within these materials is a strong spin-orbit (SO) interaction leading to a partial inversion of bands around the band gap, while respecting the time-reversal symmetry. As a consequence, non-trivial surface states emerge within the bulk energy gap, forming an odd number of Dirac cones and exhibiting a chiral relationship between spin and momentum~\cite{Fu,Moore,Roy,Fu3,Murakami}. The Landau quantization of the massless Dirac electrons in a magnetic field $B$ exhibits a square-root dependence with respect to $B$ and a field-independent zeroth Landau level (LL)~\cite{Li,Goerbig}, both, in contrast to gapped two-dimensional electron systems with parabolic dispersion. In that sense, the zeroth Landau level called LL$_0$ is a fingerprint of the topological protection of the surface states.\\
Low-temperature scanning tunneling spectroscopy (STS) can map the Landau quantization down to the atomic scale exhibiting discrete energy peaks in the differential conductivity ($dI/dV$) spectra. This method has been applied, firstly, to semiconductor systems with parabolic dispersion~\cite{Wildoer,Dombrowski,Morg, Morgenstern,Hashimoto,Hashimoto2,Becker}, and later to Dirac electron systems, such as graphene~\cite{Li,Andrei,Miller,Song}, strong 3DTIs~\cite{Cheng,Hanaguri,Okada,Jiang,Fu4}, 3D topological crystalline insulators~\cite{Okada2,Zeljikovic} and 3D Dirac semimetals~\cite{Jeon}. The method has been used to determine the energy dispersion including its lifetime broadening \cite{Wildoer,Dombrowski,Li,Andrei,Miller,Cheng,Hanaguri,Jiang,Okada2,Zeljikovic,Jeon}, to probe potential fluctuations \cite{Morg,Miller,Okada},  to probe the Landau level wave functions \cite{Hashimoto2,Fu4}, to study the related quantum Hall effect on the local scale \cite{Morgenstern,Hashimoto}, and to observe influences of electron-electron interaction \cite{Becker,Song}.

In this work, we apply the method to the phase change alloy Sb$_2$Te$_3$~\cite{Lencer}, being a prototype 3DTI with only one spin-polarized Dirac cone located at the $\overline{\Gamma}$-point, as has been revealed by spin- and angle-resolved photoemission spectroscopy (spin-ARPES)~\cite{Pauly1,Plucinski}, conventional ARPES~\cite{Wang,Seibel} and the de-Haas-van-Alphen effect~\cite{Schwartz}. Similar to other binary TIs, Sb$_2$Te$_3$ has an intrinsic defect population which pins the Fermi level $E_{\mathrm{F}}$ within the bulk valence band. Combined STM and density functional theory (DFT) calculations of Sb$_2$Te$_3$ have revealed Sb vacancies ($V_{\mathrm{Sb}}$) and Sb-on-Te antisites (Sb$_{\mathrm{Te}}$) as the energetically favorable $p$-type defects. These defects appear, e.g., after molecular beam epitaxy (MBE) of Sb$_2$Te$_3$ thin films ~\cite{Jiang1} and are responsible for the natural $p$-type conductivity~\cite{Hsieh2,Wang,Seibel}.\\
Previous STS data on MBE grown Sb$_2$Te$_3$ thin films of 7 quintuple layers (QLs)~\cite{Jiang} have found LLs corroborating the Dirac cone nature of the topological surface state, where the overall Dirac velocity was determined to be $v_{\mathrm{D}} = 4.3 \cdot 10^{5}$\,m/s. Moreover, the energy dependence of the LL width was found to be compatible with a dominating electron-electron scattering~\cite{Jiang}. Here, we show, that despite a higher defect density in bulk Sb$_2$Te$_3$ of $(2-4)\cdot 10^{12}$\,cm$^{-2}$, STS reveals Landau quantization already at a magnetic field of $B\ge 2$\,T in bulk samples. The LLs show a similar Dirac-like dispersion of the topological surface state (TSS) as found in~\cite{Jiang}. Additionally, we reveal different Dirac velocities $v_{\mathrm{D}}$ for the lower and the upper part of the Dirac cone. The distinct $v_{\mathrm{D}}$ values differ by about 20\,\%, and are both in good agreement with previous DFT calculations and ARPES data of the hole part of the Dirac cone of the same bulk crystal~\cite{Pauly1}. The Dirac point energy $E_{\mathrm{D}}$ deduced from the energy of LL$_0$ spatially varies by up to 40\,meV with respect to $E_{\mathrm{F}}$, which could be directly related to different amounts of local $p$- and $n$-type defect densities. The quasiparticle lifetime as deduced from the peak widths $\Delta E$ of the LLs reveals an approximately linear dependence of $\Delta E$ on energy with respect to $E_{\rm F}$. This is attributed to a dominant contribution of electron-electron interaction to the inelastic scattering rate by a detailed analysis of the expected features of different scattering channels, thus, generalizing the conclusion of~\cite{Jiang} towards larger defect densities and, hence, putting it on more solid grounds.

\section{Experiment}

Scanning tunneling microscopy (STM) and STS measurements are performed in a home-built microscope in ultra-high vacuum (UHV) at a temperature of $T$ = 6\,K~\cite{Mashoff} and a variable $B$ field perpendicular to the sample surface up to $B = 7$\,T. The Sb$_2$Te$_3$ single crystals are cleaved in UHV, at a base pressure of $1\cdot 10^{-10}$\,mbar, and are transferred into the precooled STM directly afterwards. STM topography images are recorded in constant-current mode at current $I$ and sample voltage $V$. The local density of states (LDOS) of the sample surface is measured via STS by locally resolved $dI/dV(V)$ curves using lock-in technique with modulation frequency $f_{\mathrm{mod}}$ = 1.5\,kHz and amplitude $V_{\mathrm{mod}}$ = 2-4\,m$V_{\mathrm{rms}}$. This results in an energy resolution of $\delta$E $\approx$ $\sqrt{(3,3k_{\mathrm{B}}T)^2 + (1,8eV_{\mathrm{mod}})^2}$ $\approx$ 4-7\,meV~\cite{Morgenstern2} as has been crosschecked for the used STM previously \cite{Mashoff}. The $dI/dV(V)$ spectra are obtained by stabilizing the tungsten tip at a distinct tip-sample distance defined by the current $I_\mathrm{stab}$ and sample voltage $V_\mathrm{stab}$.
\begin{figure}[tb]
\includegraphics[width=8.6cm]{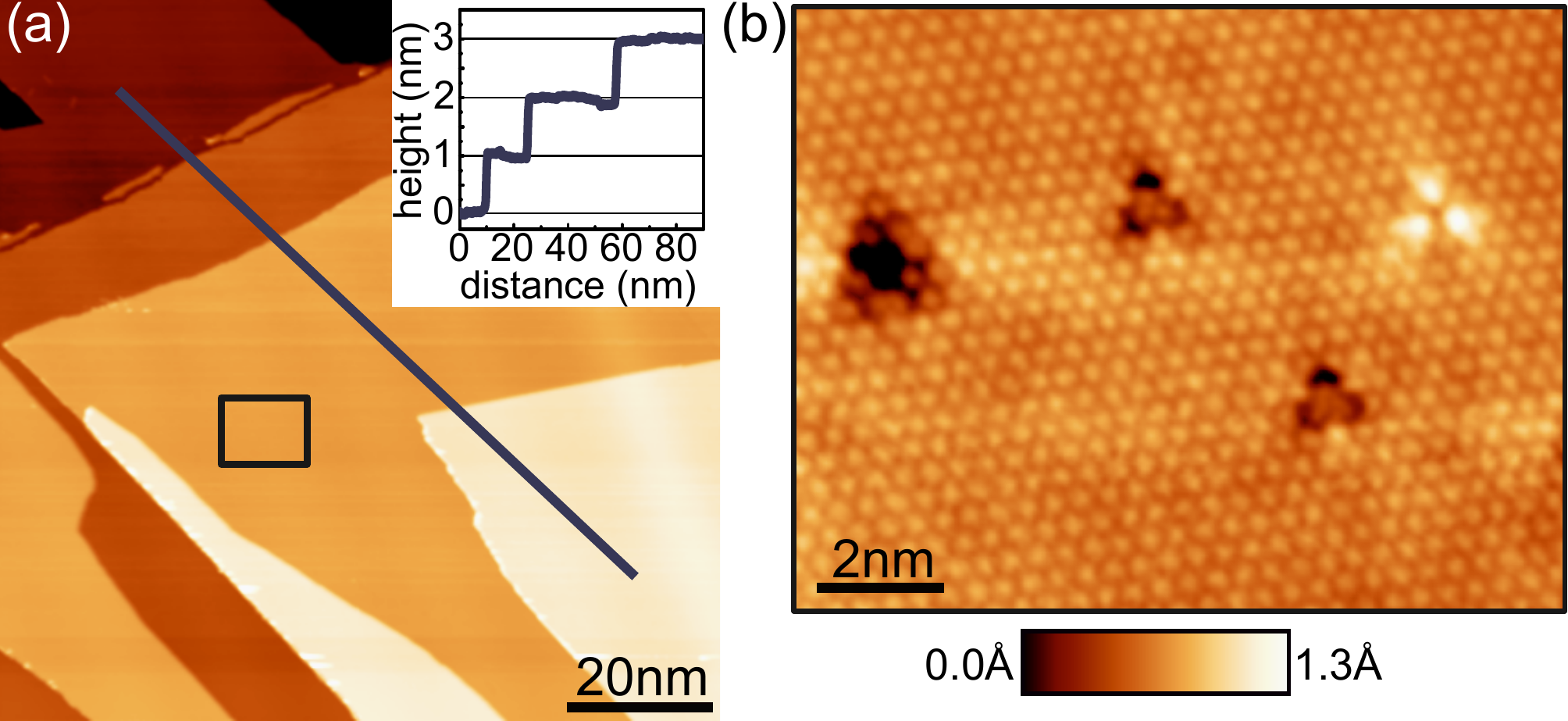}
\caption{\label{figure1}(color online) (a) Large scale STM image of cleaved Sb$_2$Te$_3$(0001) ($V=0.9$\,V, $I=50$\,pA) revealing large terraces of widths up to 50\,nm. Inset: Line profile along the blue line of the main image showing the step heights which correspond to the height of one QL ($\approx 1$\,nm). (b) Atomically resolved STM image ($V=0.4$\,V, $I=1$\,nA) recorded in the area marked by the black box in (a). A hexagonally arranged pattern of the top surface Te atoms is observed with an average atomic distance of 0.42\,nm. Different types of defects are visible as clover-shaped structures appearing dark and bright.}
\end{figure}

\section{Identification of defects}
A large-scale STM image of the cleaved Sb$_2$Te$_3$(0001) surface is shown in Fig.~\ref{figure1}(a). It reveals typical terraces with widths of up to 50\,nm separated by steps of $\approx$ 1\,nm (inset of Fig.~\ref{figure1}(a)), which corresponds to the height of one QL. Moreover, the atomic structure of the surface is visible at smaller scale (Fig.~\ref{figure1}(b)) and reveals a hexagonally arranged pattern of Te-atoms with an atomic distance of $a$ = 0.42\,nm. Typical intrinsic defects of the top QL are visible as clover-shaped darker and brighter areas in the STM image. The bright defect corresponds to a Sb-on-Te antisite (Sb$_{\mathrm{Te}}$) in the Te surface layer, while the dark defect is a Sb vacancy located in the underlying Sb-layer (V$_{\mathrm{Sb1}}$), as has both been previously identified by Jiang \textit{et al.}~\cite{Jiang1}. These particular defects have been found to be responsible for the natural $p$-type conductivity of Sb$_2$Te$_3$ \cite{Jiang1}.
\begin{figure}[tb]
\includegraphics[width=6.7cm]{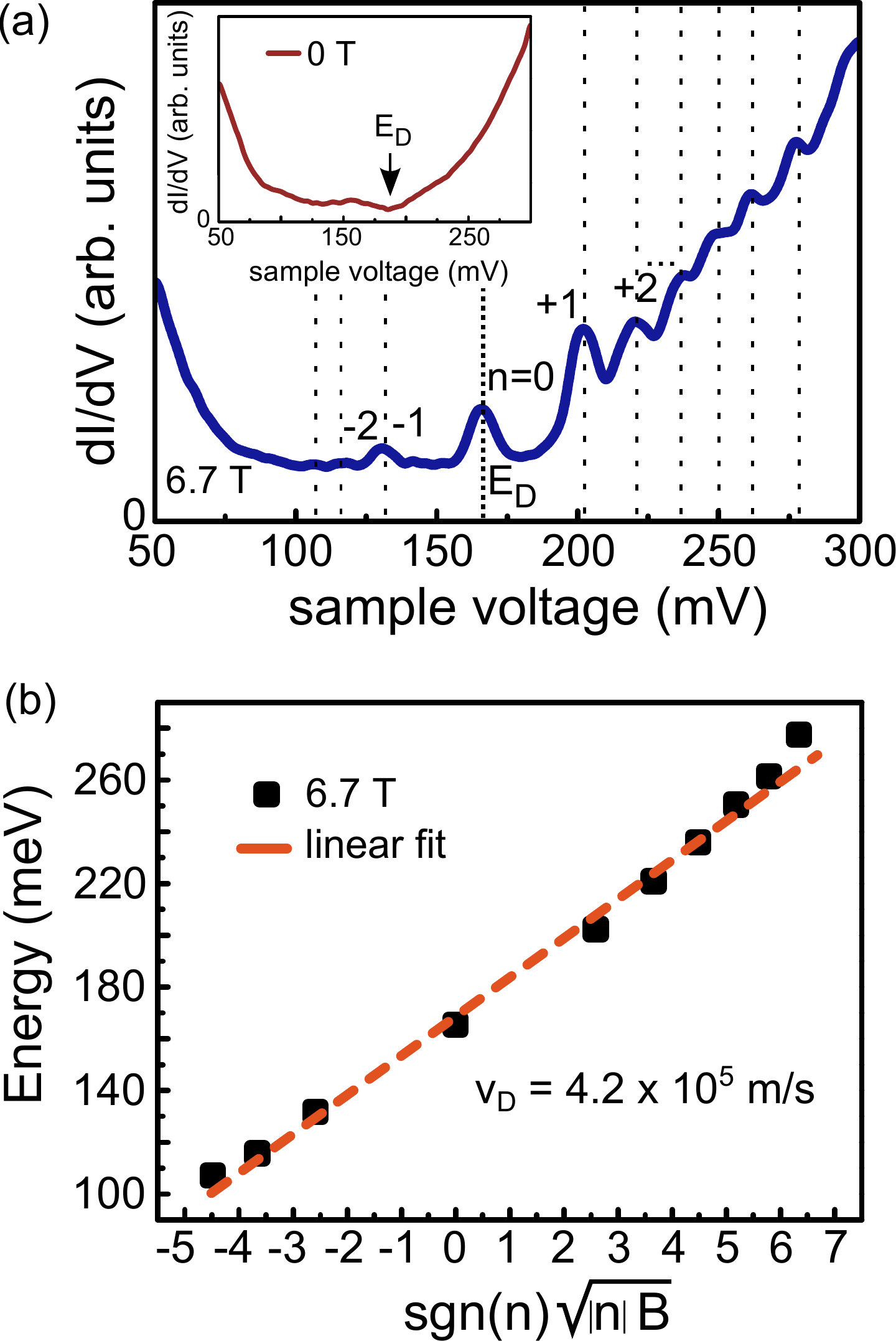}
\caption{\label{figure2}(color online) (a) $dI/dV(V)$ spectrum recorded on cleaved Sb$_2$Te$_3$(0001) at $B=6.7$\,T showing Landau quantization of the topological surface state ($V_{\rm stab}=0.3$\,V, $I_{\rm stab}=400$\,pA, $V_{\rm mod}=4$\,mV). The spectrum is an average of ten consecutive spectra measured at the same point on an area away from defects. $E_{\rm D}$ marks the position of the Dirac point located at LL$_0$, the numbers mark the LL index $n$, and the dashed lines mark the peak positions deduced from Lorentzian fits. Inset: $dI/dV(V)$ spectrum at $B=0$ T measured at the same position as the one in the main image with $E_{\rm D}$ marking the minimum of the curve ($V_{\rm stab}=0.3$\,V, $I_{\rm stab}=50$\,pA, $V_{\rm mod}=4$\,mV). (b) Landau level energies deduced from Lorentzian fits of the peaks in (a) and plotted against sgn($n$)$\sqrt{|n|B}$. The line is a linear fit resulting in the Dirac velocity $v_{\mathrm{D}}$ as indicated.}
\end{figure}

\section{Defect densities compared with Dirac point energies as deduced from LL spectroscopy}
The electronic structure of Sb$_2$Te$_3$(0001) is, firstly, probed at $B=0$ T via the $dI/dV(V)$ spectrum shown in the inset of Fig.~\ref{figure2}(a). The minimum in the differential conductivity at 170\,meV above $E_{\mathrm{F}}$ is attributed to the Dirac point energy $E_{\mathrm{D}}$ through comparison with the energy dispersion of the TSS from DFT calculation~\cite{Pauly1}. In the same way, the finite intensity for energies below the Dirac point can be attributed to the bulk valence band. Thus, the STS spectrum reveals a significant hole doping in accordance with other STM and ARPES studies~\cite{Hsieh2,Wang,Seibel,Plucinski,Jiang,Jiang1}. The related large carrier density of $2-5 \cdot 10^{19}$\,cm$^{-3}$~\cite{Takagaki, Peranio} in combination with the surface Dirac cone favorably suppresses tip induced band bending~\cite{Mashoff}. However, the found $E_{\rm D}-E_{\rm F}$ deviates from a previous ARPES study on the same Sb$_2$Te$_3$ crystal for an unknown reason~\cite{Pauly1}.

The Dirac fermion nature of the TSS can be probed by LL spectroscopy, exploiting that the electron energy $E_n$ in a perpendicular $B$-field is quantized into discrete values according to~\cite{Castro,Andrei,Cheng,Hanaguri}
\begin{equation}\label{LL}
E_n = E_{\rm D} + {\rm sgn}(n) \hbar v_{\rm D}\sqrt{2e |n| B/\hbar }, \,\,\,\,\,\,\,\,\,  n = 0, \pm1, \pm2,...
\end{equation}
with $n$ being the Landau level index. This equation includes the $B-$field independent energy of LL$_0$ at $n=0$, which is protected by the Berry phase of $\pi$ of the TSS. Figure~\ref{figure2}(a) shows the $dI/dV(V)$ spectrum measured at $B = 6.7$\,T and at an area away from defects. Note, that the LL spectra measured on top of a defect closeby are barely changed. The quantization into LLs appears as peaks with non-equal energy spacing. The latter reveals the non-parabolic dispersion of the Dirac cone. The central peak is found at about $V$ = 166\,mV, which is exactly the energy position of $E_{\mathrm{D}}$ deduced from the $dI/dV$ curve at $B=0$ T (inset of Fig.~\ref{figure2}(a)). This peak is, thus, identified as LL$_0$. Below LL$_{\mathrm{0}}$, only three further peaks are resolved probably due to the overlap of the lower part of the Dirac cone with the bulk valence band~\cite{Pauly1}. This is in qualitative agreement with a recent ARPES work, which proposes that a change in the TSS wave function occurs at the crossing point between the TSS and the bulk valence band due to hybridization~\cite{Seibel1}. The character of the TSS wave function is then expected to become more bulk-like leading to less pronounced LLs~\cite{Zhang3}. Above LL$_{\mathrm{0}}$, up to six peaks, not present at $B$ = 0\,T, are discernible.\\
\begin{figure}[tb]
\includegraphics[width=6.6cm]{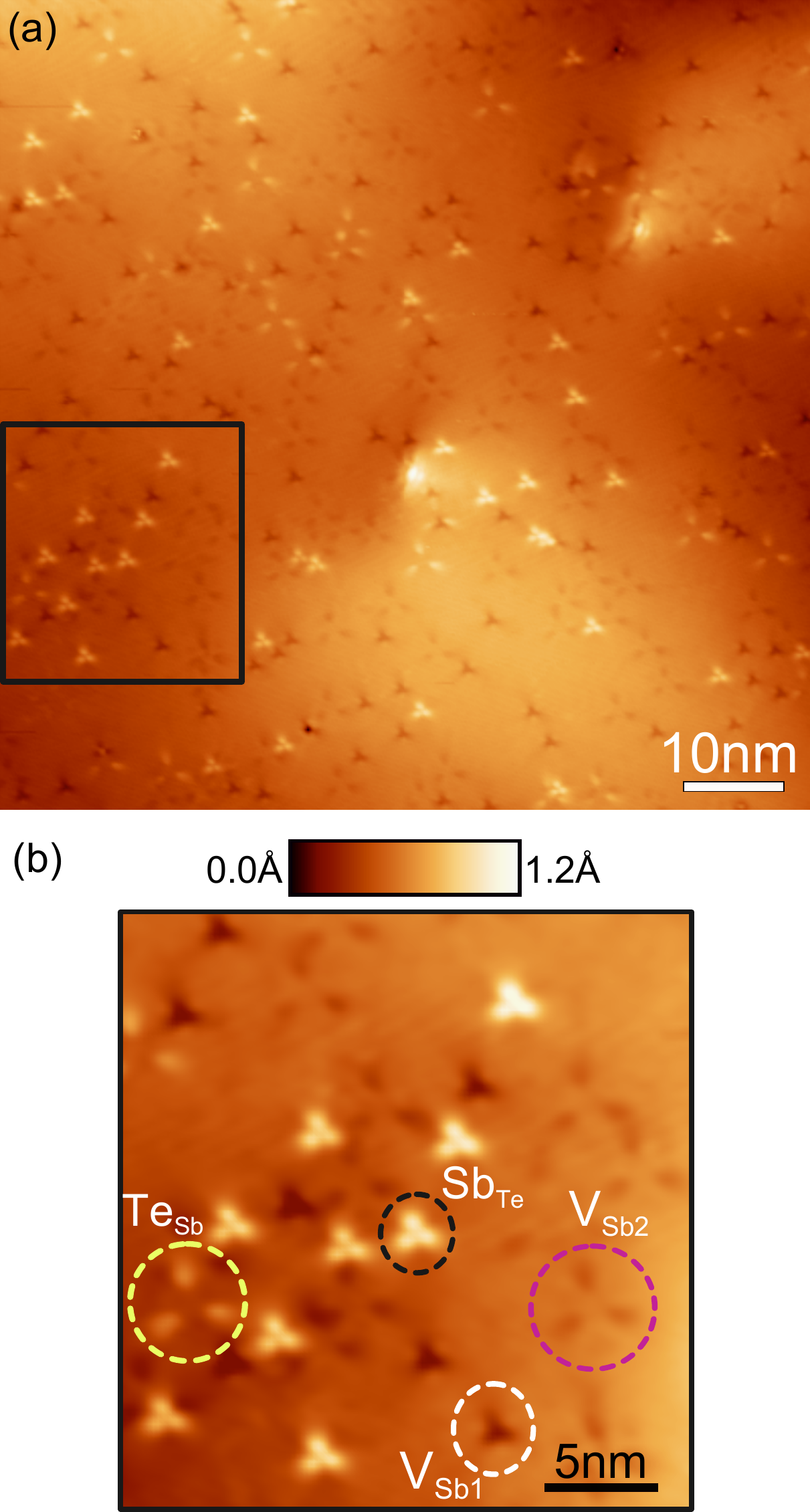}
\caption{\label{figure3}(color online) (a) STM image ($V$ = 0.9\,V, $I$ = 100\,pA) revealing typical types of defects at the surface of Sb$_2$Te$_3$(0001). (b) Zoom into the marked area of (a) showing a close up view with four different types of defects marked by different colored ellipses, which are labeled according to \cite{Jiang1}.}
\end{figure}
The dispersion of the probed electron states can be deduced by plotting the found LL energies $E_{\mathrm{n}}$ with respect to sgn($n$)$\sqrt{|n|B}$ (eq.~\ref{LL}). Therefore, the $dI/dV$ curve of Fig. \ref{figure2}(a) is fitted by a Lorentzian function for each peak leaving the peak positions and the peak widths as free parameters. The resulting peak positions $E_n$ are indicated by dashed lines in Fig. \ref{figure2}(a). The respective plot $E_n({\rm sgn}(n)\sqrt{|n|B})$ is shown in Fig.~\ref{figure2}(b). For the Landau levels around LL$_{0}$, it reveals the expected linear dispersion close to $E_{\mathrm{D}}$ confirming the massless Dirac fermion character of the TSS. Towards higher LL index ($n$ $\ge$ 4, $n$ $\leq$ -3), the dispersion gets slightly non-linear. A non-linearity away from $E_{\mathrm{D}}$ has also been found in previous DFT calculations of Sb$_2$Te$_3$(0001) ~\cite{Pauly1}. The resulting $v_{\rm D}$ deduced from a linear fit to the data amounts to $v_{\mathrm{D}} = 4.2 \pm 0.1 \cdot 10^{5}$\,m/s, which nicely agrees with the value obtained by LL spectroscopy on a 7 QL film of Sb$_2$Te$_3$ grown by MBE ($v_{\mathrm{D}} = 4.3 \cdot 10^{5}$\,m/s)~\cite{Jiang}.\\
\begin{table*}
\begin{ruledtabular}
\centering
\caption{Number of specific $p$- and $n$-type defects (as marked in Fig.~\ref{figure3}(b)) found in six distinct areas, all being 40 $\times$ 40\,nm$^2$ in size. The counting uncertainty amounts to 10\,\%. The overall defect density $n_{\rm D}$ is the sum of all defects divided by the probed area. The effective $p$-type doping $p_{\rm eff}$ is the sum of the $p$-type defects minus the number of the $n$-type defects divided by the product of probed area and thickness of a QL (1 nm). The energy position of the Dirac point $E_{\mathrm{D}}$ for area 1 and 2 is deduced from the fitted position of LL$_{0}$. Area 1 (Area 2) is the region of the LLs shown in Fig.~\ref{figure2}(a) (Fig.~\ref{figure4}(a)).}
\label{tab1}
\begin{tabular}{|cl|c|c|c|c|c|c|}
\hline
\multicolumn{2}{|l|}{} & Area 1 & Area 2 & Area 3 & Area 4 & Area 5 & Area 6 \\
                      & & (LLs in Fig.~\ref{figure2}(a)) & (LLs in Fig.~\ref{figure4}(a)) & (area within Fig.~\ref{figure3}(a)) & & &  \\ \hline
\multirow{2}{*}{$p$-type defects:}  & V$_{\mathrm{Sb}}$ & 32 & 56 & 48 & 53 & 49 & 32 \\ 
                   & Sb$_{\mathrm{Te}}$  & 14 & 6 & 10 & 3 & 3 & 8 \\ 
                  $n$-type defects: & Te$_{\mathrm{Sb}}$ & 3 & 3 & 6 & 2 & 4 & 2 \\ \hline
\multicolumn{2}{|c|}{overall defect density} & 0.031\,nm$^{-2}$ & 0.041\,nm$^{-2}$ & 0.040\,nm$^{-2}$ & 0.036\,nm$^{-2}$ & 0.035\,nm$^{-2}$ & 0.026\,nm$^{-2}$ \\ \hline
\multicolumn{2}{|c|}{effective $p$-type doping} & 2.7 $\cdot$ 10$^{19}$\,cm$^{-3}$ & 3.7 $\cdot$ 10$^{19}$\,cm$^{-3}$ & 3.2 $\cdot$ 10$^{19}$\,cm$^{-3}$ & 3.4 $\cdot$ 10$^{19}$\,cm$^{-3}$ & 3.0 $\cdot$ 10$^{19}$\,cm$^{-3}$ & 2.4 $\cdot$ 10$^{19}$\,cm$^{-3}$ \\ \hline
\multicolumn{2}{|c|}{$E_{\mathrm{D}}-E_{\rm F}$} & $(166.1 \pm 0.05)$\,meV & $(204\pm 1)$\,meV & / & / & / & / \\ \hline
\end{tabular}
\end{ruledtabular}
\end{table*}

Next, we determine the local defect density within the top QL.  This defect density will locally determine the electrostatic potential as measurable by $E_{\rm D}-E_{\rm F}$ with $E_{\rm D}$ being the energy of LL$_{0}$. We rely on the comparative STM and DFT study mentioned above ~\cite{Jiang}. This study has identified the Sb vacancies (V$_{\mathrm{Sb}}n$) in the two different Sb-layers within the top QL ($n=1,2$: layer label), and the Sb$_{\mathrm{Te}}$ antisites located at the top Te-layer. These defects are acceptors, while the also identified Te-on-Sb antisites (Te$_{\mathrm{Sb}}$) act as donors. Note, that Sb$_{\mathrm{Te}}$ antisites on subsurface Te layers of the top QL are not identified within our STM data, even though they should be visible with reduced intensity at the surface, most probably due to their higher formation energy~\cite{Jiang}.\\
Figure~\ref{figure3}(a) shows a STM image of the Sb$_2$Te$_3$(0001) surface, resolving different types and numbers of clover-shaped defects. The apparent defect density is  $3.9 \cdot 10^{12}$\,cm$^{-2}$, which is slightly larger than defect densities found in MBE grown thin films of Sb$_2$Te$_3$~\cite{Jiang1} ($1.7 \times 10^{12}$\,cm$^{-2}$) and Bi$_2$Se$_3$~\cite{Cheng} ($\sim$ 10$^{12}$\,cm$^{-2}$). The defects are labeled by comparison with the previous results ~\cite{Jiang1} within the close-up picture of Fig.~\ref{figure3}(b). The two types of Sb vacancies (V$_{\mathrm{Sb1}}$, V$_{\mathrm{Sb2}}$) exhibit depressions of different lateral size at positive sample voltage, while the two antisite defects (Sb$_{\mathrm{Te}}$, Te$_{\rm Sb}$) appear as protrusions of different lateral size. Obviously, the acceptors (V$_{\mathrm{Sb1}}$, V$_{\mathrm{Sb2}}$, Sb$_{\mathrm{Te}}$) outnumber the donors (Te$_{\rm Sb}$) by far as found consistently on all probed areas of the sample. This is in line with the known $p$-type conductivity of Sb$_2$Te$_3$.\\
To this end, we used the identification of different defects in order to determine defect densities for several 40 $\times$ 40\,nm$^2$ surface areas as summarized in table~\ref{tab1}. For two of these areas, we additionally performed LL spectroscopy in order to determine $E_{\rm D}-E_{\rm F}$. The LL spectroscopy recorded at different spatial positions and a single field (6.7\,T) within the 40 $\times$ 40\,nm$^2$ areas revealed that $E_{\rm D}$ shifts by less than 5 meV, which points to more long range potential fluctuations. Note, however, that the spatial sensitivity of LL$_0$ for potential fluctuations is limited by the magnetic length $l_B=\sqrt{\hbar/(eB)}$ being $l_B = 10$ nm at $B=6.7$ T \cite{Hashimoto}. This also prohibits to detect potential fluctuations caused by individual defects, since the LL wave functions always cover several defects.\\
From the counted amounts of different defects, we determined straightforwardly the overall defect density $n_{\rm D}$ by dividing the sum of all defects by the probed area. A first estimate of the effective $p$-type doping $p_{\rm eff}$ results from subtracting the number of $n$-type defects from the sum of the $p$-type defects and dividing by the product of probed area and thickness of a QL (1\,nm). This assumes singly charged defects due to the fact that their charge is not known~\cite{Jiang1} and cannot be determined by STS. The numbers for $E_{\rm D}$, $n_{\rm D}$, and $p_{\rm eff}$ are also displayed in table \ref{tab1} revealing, e.g., average values of $\overline{n}_{\rm D} = 0.035$\,nm$^{-2}$ and $\overline{p}_{\rm eff} = 3.1 \times 10^{19}$\,cm$^{-3}$. The hole concentration of Sb$_2$Te$_3$ has previously been deduced from transport measurements to be $5.3 \cdot 10^{19}$\,cm$^{-3}$ for thin films grown by MBE~\cite{Takagaki} and to be $2.7 \cdot 10^{19}$\,cm$^{-3}$ for a polycrystal ~\cite{Peranio}, both fits reasonably with the effective $p$-type doping deduced from our STM images.\\
By comparing area 1 and area 2,  one observes that an increased $p_{\rm eff}$ leads to a larger value of $E_{\rm D}-E_{\rm F}$ as expected. Moreover, the increase of $E_{\rm D}-E_{\rm F}$ ($\sim 25$ \%) is smaller than the increase in $p$-type doping ($\sim 40$ \%) indicating an increasing density of states at $E_{\rm F}$ with increasing distance from $E_{\rm D}$, as expected for the Dirac cone. Quantitatively, the expected charge carrier density of the Dirac cone of $n(E)=(E-E_{\rm D})^2/(4\pi\hbar^2 v_{\rm D}^2)$ amounts to $n$(166 meV) = $2.8 \cdot 10^{12}$\,cm$^{-2}$, respectively, $n$(204 meV) = $4.2 \cdot 10^{12}$\,cm$^{-2}$, which nicely fits to the defect density observed by STM (table \ref{tab1}) when assuming one single charge per defect. However, this excellent agreement is probably coincidental, since, on the one hand, STM is only sensitive to the first QL \cite{Jiang}, while DFT calculations show that the hole part of the TSS penetrates by more than one QL into the bulk of the substrate \cite{Pauly1,Zhang3}, which increases the effective doping with respect to the one deduced from the counting of defects in STM measurements. On the other hand, the lower part of the TSS energetically overlaps with bulk states also to be doped by defects, which effectively decreases the achievable $E_{\rm D}-E_{\rm F}$ at a given dopant density.\\
The fluctuation of the number of defects in different areas of $40 \times 40$\,nm$^2$ is remarkably large. For V$_{\rm Sb}$, we find an average number of $\overline{N}_{\rm V_{\rm Sb}}=45$ with a standard deviation of $\sigma_{\rm V_{\rm Sb}}=10.5$, while the statistical fluctuation should be $\overline{\sigma}=6.7$ only. The deviation between $\sigma_{\rm V_{\rm Sb}}$ and $\overline{\sigma}$ by nearly 60 \% is significant regarding the six sample areas probed. The same trend, but without statistical significance, is also found for the Sb$_{\rm Te}$ defects. Thus, we believe that the potential fluctuations are not only given by a statistical distribution of defects, but are also influenced by the kinetics of defect formation depending critically on local temperature \cite{Jiang} and local flux of the different constituents. This implies stronger potential fluctuations on large length scales as anticipated by a statistical distribution of acceptors only \cite{Shklovskii}.\\
\begin{figure*}[htb]
\includegraphics[angle=0,width=15cm]{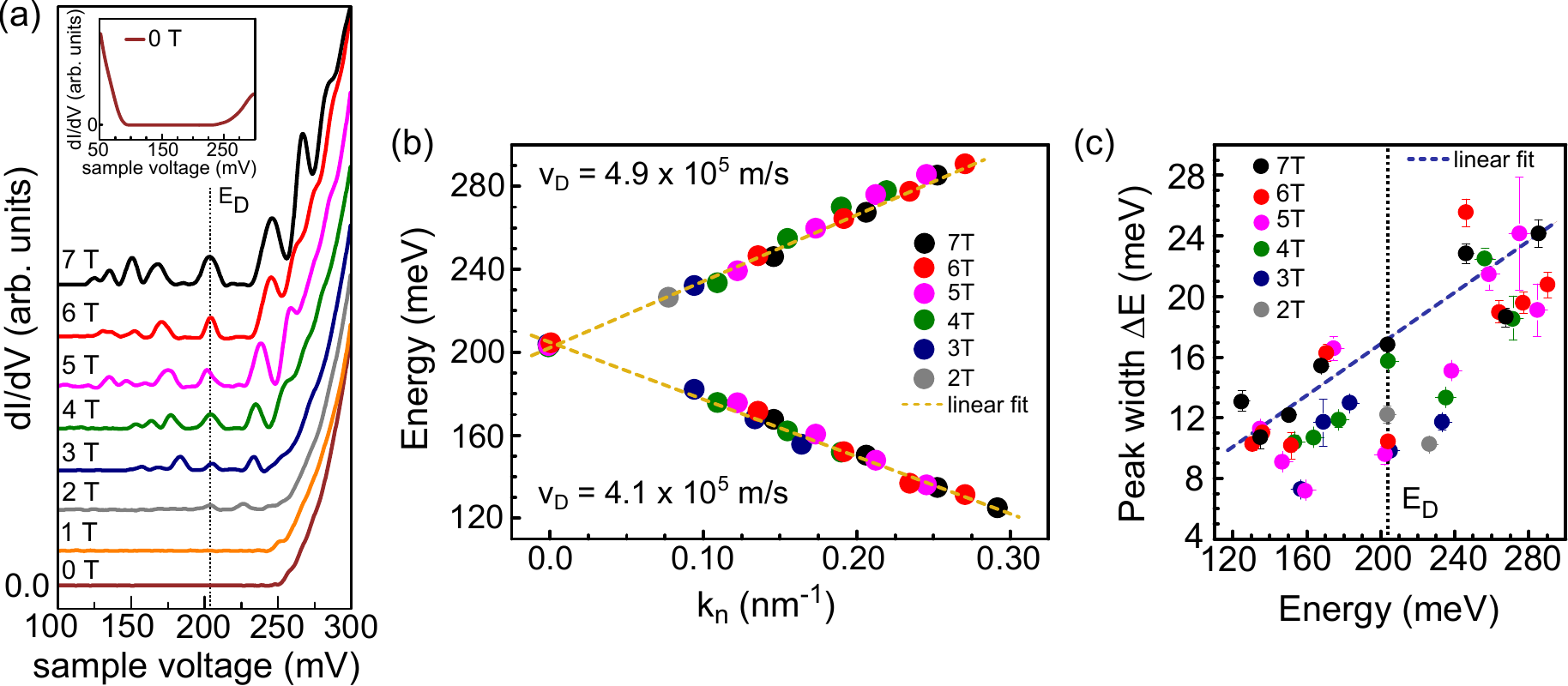}
\caption{\label{figure4}(color online) (a) $dI/dV(V)$ spectra recorded at $B = 0 -7$\,T as marked on a different sample area as the LLs from Fig.~\ref{figure2}(a) ($V_{\mathrm{stab}}=0.3$\,V, $I_{\mathrm{stab}}=100$\,pA, $V_{\mathrm{mod}}=2$\,mV). For direct comparison of the spectra, each spectrum (at a distinct field), which is again an average of 10 spectra, is measured on the same point away from the defects. The vertical dotted line indicates the field-independent zeroth Landau level at the Dirac point $E_{\rm D}$. The spectra are shifted vertically for clarity. Inset: $dI/dV(V)$ spectrum for a larger energy range  measured at $B=0$ T in the same sample area ($V_{\rm stab}=0.3$\,V, $I_{\rm stab}=100$\,pA, $V_{\rm mod}=2$\,mV). (b) Landau level energies deduced from Lorentzian fits of the curves in (a) and plotted versus momentum $k_n$ as deduced from eq.~\ref{momentum}. The dashed lines are linear fits to the lower and upper part, respectively, with resulting Dirac velocities $v_{\mathrm{D}}$ marked. (c) Full width at half maximum for each Landau level, as deduced by a Lorentzian fit function, plotted as a function of energy. Different $B$ fields are indicated. The dashed line is a linear fit forced through $\Delta E=0$ meV at $E_{\rm F}$.}
\end{figure*}

\section{Different Dirac velocities for holes and electrons}

Figure~\ref{figure4}(a) shows the LL measurement for area 2 at varying $B$. At $B>2$\,T, quantization peaks are visible, which become more pronounced with increasing $B$. The field-independent LL$_{\mathrm{0}}$, identified as the Dirac point $E_{\rm D}$, is found at $E_{\mathrm{D}}$ = 204\,meV (see table \ref{tab1}). Its field-independence corroborates again the Dirac fermion nature of the TSS. In area 2, $E_{\rm D}$ could not be identified at $B=0$\,T due to a low $dI/dV$ intensity of the TSS within the band gap (inset in Fig.~\ref{figure4}(a)), which is most probably due to a different density of states of the microtip in comparison with the microtip used to probe area 1 (inset of Fig.~\ref{figure2}(a)). This underlines the advantage of LL spectroscopy to determine $E_{\rm D}$ precisely.\\
Figure~\ref{figure4}(b) shows the derived Landau level energies $E_n$, again deduced from the peak values of Lorentzian functions fitting the $dI/dV$ curves. They are displayed with respect to their momentum $k_n$ as derived from the Onsager semiclassical quantization condition~\cite{Roth}. This condition provides an equation for the $k$-space area $S_n$ of the $n$th Landau orbits, namely $S_n$ = 2$\pi$($n$+$\gamma$)(1/$l_{B}^{2}$), with $\gamma$ = 0 being the phase offset for Dirac fermions~\cite{Wright}. Close to $E_{\mathrm{D}}$, the Dirac cone in Sb$_2$Te$_3$ is isotropic~\cite{Pauly1} such that $S_n$ = $\pi$$k_n^2$ leads to:
\begin{equation}\label{momentum}
k_n = \sqrt{\frac{2e\left|n\right|B}{\hbar}}.
\end{equation}
The corresponding $E_{n}(k_{n})$ exhibits the known linear band dispersion, however, with different $v_{\rm D}$ for the lower and the upper part of the Dirac cone, namely $v_{\mathrm{D}} = 4.1\pm 0.1 \cdot 10^{5}$\,m/s and $v_{\mathrm{D}} = 4.9\pm 0.2 \cdot 10^{5}$\,m/s, respectively. A slower $v_{\mathrm{D}}$ below $E_{\mathrm{D}}$ has also been found for the 3DTI Bi$_2$Se$_3$~\cite{Xia}. The value for the lower part of Sb$_2$Te$_3$ (0001) reasonably fits with ARPES measurements of the same crystal ($v_{\mathrm{D}} = 3.8\pm 0.2 \cdot 10^{5}$\,m/s)~\cite{Pauly1}. The results also roughly agree with the $v_{\rm D}$ deduced from DFT calculations, which reveal $v_{\mathrm{D}} = 3.3 \cdot 10^{5}$\,m/s for the lower and $v_{\mathrm{D}} = 5.0 \cdot 10^{5}$\,m/s for the upper part of the Dirac cone within the same momentum range~\cite{Pauly1}. Note, that a $v_{\mathrm{D}}$ of $4.4 \pm 0.1 \cdot 10^{5}$\,m/s is obtained if evaluated in the same way as in Fig.~\ref{figure2}(b), which nicely agrees with $v_{\mathrm{D}} = 4.2 \pm 0.1 \cdot 10^{5}$\,m/s deduced in Fig.~\ref{figure2}(b) and $v_{\mathrm{D}} = 4.3 \cdot 10^{5}$\,m/s found in \cite{Jiang}.\\

Notice, that the area encircled by a full cyclotron orbit, as semiclassically required for Landau quantization \cite{Roth}, contains a significant amount of defects $N_{\rm D}$. For example, we observe the LL with $n=1$ on the electron side at $B=2$ T (Fig. \ref{figure4}(a)) indicating the presence of a fully developed cyclotron orbit. Taking the cyclotron radius $r_{\rm c}=\sqrt{2n+1}\cdot l_B =32$ nm, we get $N_{\rm D}=\pi\cdot r_{\rm c}^2\cdot n_{\rm D}=130$ defects. DFT shows that the penetration depth of the TSS on the electron side is about 1 QL \cite{Pauly1}, such that this number is indeed a good estimate for the number of defects encircled. In order to set this apparently large number into perspective, we estimate the cross section for electron-defect scattering classically assuming the full cyclotron orbit and get $\sigma_{\rm 2D} \le (\sqrt{2}\cdot 2\pi r_{\rm c}\cdot n_{\rm D} )^{-1}=0.8$ ${\rm \AA}$ in 2D, or regarding the depth of the QL (1 nm) a reasonable 3D cross section of $\sigma_{\rm 3D} \le 8$ ${\rm \AA}^2$.

\section{Quasiparticle lifetimes}

Finally, the energy dependence of the quasiparticle lifetime is extracted from the peak widths $\Delta E$ of the LLs as deduced from the Lorentzian fits. The result is shown in Fig.~\ref{figure4}(c) revealing a nearly linear increase of $\Delta E$  with energy. Recall that we operate at an energy resolution $\delta E = 4$ meV being smaller than the determined peak widths. A linear fit to the data in Fig. \ref{figure4}(b), forced to $\Delta E=0$ meV at $E_{\rm F}$, results in $\Delta E= (0.084\pm 0.003) \cdot (E-E_{\rm F})$ (dashed line). In addition, one could anticipate a slight decrease of $\Delta E$ around $E_{\rm D}$. We will offer a different interpretation of the apparent jump at $E \simeq$ 250\,meV later, but would like to stress that the linear behavior below $E_{\rm D}$ remains, even when excluding the data above $E_{\rm D}$.
The dip and the linear slope was observed similarly by LL spectroscopy on Sb$_2$Te$_3$ thin films \cite{Jiang} and on Bi$_2$Se$_3$ bulk material cleaved in-situ \cite{Hanaguri}. Interestingly, the linear slope of $\Delta E(E)$ towards $E_{\rm F}$ is nearly identical for the Sb$_2$Te$_3$ thin film, if applied for $E-E_{\rm F} > 50$ meV ($\simeq 0.085$) \cite{Jiang}, and only about 30 \% smaller for the Bi$_2$Se$_3$ bulk crystal ($\simeq 0.06$) \cite{Hanaguri}, albeit, in both cases, the dip at $E_{\rm D}$ is more pronounced. This is even more remarkable, since the defect density of the Sb$_2$Te$_3$ thin film is a factor of two lower than the defect density $\overline{n}_{\rm D}$ of the bulk crystal studied here.\\
The peak width being larger than $\delta E$ could be firstly caused by disorder broadening. Basically, electronic states in $B$ field have a width of about $r_{\rm c}$ and are mostly localized by the disorder \cite{Ando,Morgenstern}. Consequently, states with different origin in space and, thus, different potential energy spatially overlap resulting in a finite width of the peak in $dI/dV$, which, in first order, depends linearly on the product of potential gradient and $r_{\rm c}$ \cite{Champel}.
Since $r_{\rm c}$ increases with Landau level number $n$, the disorder broadening must also increase with $|n|$, which is in contrast to the experimental result on the hole side of $E_{\rm D}$. However, it might account for the dip of $\Delta E(E)$ at $E_{\rm D}$. The effect is most likely less pronounced in our experiment, since we operate at a larger $E_{\rm D}-E_{\rm F}$ than for the thin films \cite{Hanaguri}, such that the linear contribution of $\Delta E(E_{\rm D}-E_{\rm F})$, being of different origin, is already larger. In line, within areas of about $40 \times 40$ nm$^2$, ($r_{\rm c}=40$ nm, e.g. at $B= 3.5$ T and $n=4$), the peak position of LL$_0$ does not shift by more than 5 meV, which is on the order of the scattering of $\Delta E$, showing that disorder broadening is not the dominant effect for $\Delta E$.\\
Thus, we attribute the peak width to inelastic scattering, which could be either driven by electron-electron or by electron-phonon interaction.
The energy dependence of electron-phonon interaction is typically dominated by the available phase space for electron scattering \cite{Grimvall,Echenique}, at least, at energies above the optical phonon energy being $E_{\rm LO}=20$ meV for Sb$_2$Te$_3$ \cite{Sosso,Richter}. For LL distances smaller than $E_{\rm LO}$, one could anticipate the Dirac cone density of states $D(E, B=0\hspace{1mm} {\rm T})\propto |E-E_{\rm D}|$ as the available phase space, while for larger LL separations the scattering should be restricted to one Landau level exhibiting a degeneracy $D_{\rm LL}(E, B)\propto B$. Thus, one would expect $\Delta E$ to increase with $|n|$, i.e., away from $E_{\rm D}$ for $|n|\ge 1$ and for the well separated
LL${_0}$, one would expect $\Delta E \propto B$.
At energies below $E_{\rm D}$, bulk states start to overlap with the Dirac cone \cite{Pauly1,Wang,Seibel,Plucinski} increasing the phase space for electron scattering even further, such that one would expect shorter lifetimes on the hole side than on the electron side in obvious discrepancy to the experiment. Also the other trends anticipated for electron-phonon scattering are not observed experimentally.\\
Consequently, by the exclusion principle, we believe that electron-electron
scattering, which typically implies an increasing scattering rate with distance from $E_{\rm F}$ \cite{Burgi} is dominating $\Delta E$. Indeed, at $B=0$ T, one expects, a linear energy dependence for a Dirac cone with $E_{\rm D}=E_{\rm F}$ \cite{Gonzales} or for a doped Dirac cone at $E\ge E_{\rm D}-E_{\rm F}$ being interpolated by a more quadratic dependence towards $E_{\rm F}$ \cite{Hwang}. The numbers calculated for graphene and graphite in these studies \cite{Gonzales,Hwang} are surprisingly similar to the ones we observe for Sb$_2$Te$_3$(0001), e.g., being $\Delta E =20$ meV at $E-E_{\rm F}=200$ meV \cite{Gonzales}.  However, this might be a coincidence, since a quantitative comparison would require a more detailed calculation including the $B$ field, the penetration depth of the surface state into the bulk, which effectively changes the dielectric constant, and the bulk valence bands as additional scattering channels. Notice that the peak widths jump from about 15\,meV to about 22\,meV at an energy of $\simeq$ 250\,meV. This coincides with the onset of an increasing $dI/dV$ signal (Fig.~\ref{figure4}(a)) most likely being related to the onset of the bulk conduction band. This onset is also found in two-photon ARPES data to be about 100\,meV above the Dirac point \cite{Sanchez2}. Thus, interband scattering sets in as soon as the Landau levels from the Dirac cone overlap with the bulk conduction band.
The fact that time-resolved ARPES data at $B=0$ T find relaxation times of about $\tau =1$ ps at $E_{\rm D}-E_{\rm F} = 0-100$ meV \cite{Sanchez2,Reimann}, which would correspond to $\Delta E=\hbar/\tau =0.7$ meV indicates that either the $B$ field and the related localization of electrons or the presence of the STM tip might change the scattering rates significantly.\\
Notice that the lifetimes observed by the fit of the line widths of ARPES data \cite{Sanchez,Chen} are typically even shorter than the ones observed by
LL spectroscopy \cite{Hanaguri,Jiang} and, thus, in stronger discrepancy to the time-resolved data \cite{Hajlaoui,Crepaldi}. This is most likely due to disorder averaging, i.e. bands from areas with different $E_{\rm D}$  (see table \ref{tab1}) overlap in the $E(k)$ representation of the ARPES data. Notice further that a dominating influence of electron-electron interaction as a relaxation channel of hot electrons has also been deduced from the decay of standing waves from step edges of Bi$_2$Se$_3$(0001) thin films probed by STM at $B=0$ T \cite{Song2}. The authors deduced that $\tau$ decreases with $|E-E_{\rm F}|$ finding $\tau =100-10$ fs at $|E-E_{\rm F}|=0.05-0.3$ eV.\\

\section{Summary}
In summary, we probed the Landau quantization of the topological surface state of Sb$_2$Te$_3$(0001) by STS for varying magnetic fields perpendicular to the sample surface. Our data reveal the Dirac type energy dependence, $E_n$ $\propto$ $\sqrt{|n|B}$, with different Dirac velocities for the hole and the electron branch. The Dirac point energy deduced from the zeroth Landau level spatially fluctuates by about 40 meV which could be traced back to varying densities of $p$-type defects beyond a statistical distribution. The peak width of the Landau levels is found to be $\Delta E \simeq 0.08\cdot |E-E_{\rm F}|$ pointing to a  dominating influence of electron-electron interaction.\\

\section{Acknowledgment}
We gratefully acknowledge provision of the sample by M. Wuttig, technical help by M. Pratzer, and financial support by the DFG via SFB 917, project A3 and Mo 858/13-1 as well as of Fonds National de la Recherche (Luxembourg).



\begin{thebibliography}{99}
\bibitem{Kane}
C.~L.~Kane and E.~J.~Mele, Phys.~Rev.~Lett. {\bf 95}, 146802 (2005).
\bibitem{Schnyder}
A.~P.~Schnyder, S.~Ryu, A.~Furusaki and A.~W.~W. Ludwig, Phys.~Rev.~B {\bf 78}, 195125 (2008).
\bibitem{Hasan}
M.~Z.~Hasan and C.~L.~Kane, Rev.~Mod.~Phys. {\bf 82}, 3045 (2010).
\bibitem{Qi}
X.~L.~Qi and S.~C.~Zhang, Rev.~Mod.~Phys. {\bf 83}, 1057 (2011).
\bibitem{Yan}
B.~Yan and S.~C.~Zhang, Rep.~Prog.~Phys. {\bf 75}, 096501 (2012).
\bibitem{Thouless}
D.~J.~Thouless, M.~Kohmoto, M.~P.~Nightingale and M.~den Nijs, Phys.~Rev.~Lett. {\bf 49}, 405 (1982).
\bibitem{Kohmoto}
M.~Kohmoto, Ann.~Phys. {\bf 160}, 343 (1985).
\bibitem{Volovik}
G. E. Volovik, Sov. Phys. JETP {\bf 67}, 1804 (1988).
\bibitem{Konig}
M.~K\"onig, S.~Wiedmann, C.~Br\"une, A.~ Roth, H.~Buhmann, L.~W.~Molenkamp, X-L.~Qi, and S-C.~Zhang, Science {\bf 318}, 766 (2007).
\bibitem{Volkov}
B. A. Volkov and O. A. Pankratov, JETP Lett. {\bf 42}, 145 (1985).
\bibitem{Kane2}
C.~L.~Kane and E.~J.~Mele, Phys.~Rev.~Lett. {\bf 95}, 226801 (2005).
\bibitem{Fu}
L.~Fu, C.~L.~Kane and E.~J.~Mele, Phys.~Rev.~Lett. {\bf 98}, 106803 (2007).
\bibitem{Hsieh}
D.~Hsieh {\it et al.}, Science {\bf 323}, 919 (2009).
\bibitem{Zhang}
H.~Zhang, C-X.~Liu, X-L.~Qi, X.~Dai, Z.~Fang and S-C.~Zhang, Nature Phys. {\bf 5}, 438 (2009).
\bibitem{Ringel}
Z.~Ringel, Y.~E.~Kraus, and A.~Stern, Phys.~Rev.~B {\bf 86}, 045102 (2012).
\bibitem{Rasche}
B.~Rasche, A.~Isaeva, M.~Ruck, S.~Borisenko, V.~Zabolotnyy, B.~B\"uchner, K.~Koepernik, C.~Ortix, M.~Richter, and J.~ van den Brink, Nature Mat. {\bf 12}, 422 (2013).
\bibitem{Pauly}
C.~Pauly, B.~Rasche, K.~Koepernik, M.~Liebmann, M.~Pratzer, J.~Kellner, M.~Eschbach, B.~Kaufmann, L.~Plucinski, C.~M.~Schneider, M.~Ruck, J.~van den Brink, and M.~Morgenstern, Nature Phys. {\bf 11}, 338 (2015).
\bibitem{Fu2}
L.~Fu, Phys.~Rev.~Lett. {\bf 106}, 106802 (2011).
\bibitem{Dziawa}
P.~Dziawa, B.~J.~Kowalski, K.~Dybko, R.~Buczko, A.~Szczerbakow, M.~Szot, E.~Lusakowska, T.~Balasubramaniam, B.~M.~Wojek, M.~H.~Berntsen, O.~Tjernberg, and T.~Story, Nature Mat. {\bf 11}, 1023 (2012).
\bibitem{Yu}
R.~Yu, W.~Zhang, H-J.~Zhang, S-C.~Zhang, X.~Dai, Z.~Fang, Science {\bf 329}, 61 (2010).
\bibitem{Chang}
C.~Z.~Chang, J.~Zhang, X.~Feng, J.~Shen, Z.~Zhang, M.~Guo, K.~Li, Y.~Ou, P.~Wei, L-L.~Wang, Z-Q.~Ji, Y.~Feng, S.~Ji, X.~Chen, J.~Jia, X.~Dai, Z.~Fang, S-C.~Zhang, K.~He, Y.~Wang, L.~Lu, X-C.~Ma, and Q-K.~Xue, Science {\bf 340}, 167 (2013).
\bibitem{Zhang2}
X.~Zhang, H.~Zhang J.~Wang, C.~Felser, and S-C.~Zahn, Science {\bf 335}, 1464 (2012).
\bibitem{Yang}
K.~Yang, W.~Setyawan, S.~Wang, M.~Buongiorno Nardelli, and S.~Curtarolo, Nature Mat. {\bf 11}, 614 (2012).
\bibitem{Moore}
J.~E.~Moore and L.~Balents, Phys.~Rev.~B {\bf 75}, 121306 (2007).
\bibitem{Roy}
R.~Roy, Phys.~Rev.~B {\bf 79}, 195322 (2009).
\bibitem{Fu3}
L.~Fu and C.~L.~Kane, Phys.~Rev.~B {\bf 76}, 045302 (2007).
\bibitem{Murakami}
S.~Murakami, New. J. Phys. {\bf 9}, 356 (2007).
\bibitem{Li}
G.~Li and E.~Y.~Andrei, Nature Phys. {\bf 3}, 623 (2007).
\bibitem{Goerbig}
M.~O.~Goerbig, Rev.~Mod.~Phys. {\bf 83}, 1193 (2011).
\bibitem{Wildoer}
J.~W.~G.~Wild\"oer, C.~J.~P.~M.~Harmans, and H.~van Kempen, Phys.~Rev.~B {\bf 55}, 16013 (1997).
\bibitem{Dombrowski}
R.~Dombrowski, C.~Wittneven, M.~Morgenstern, and R.~Wiesendanger, Appl. Phys. A {\bf 66}, 203 (1998).
\bibitem{Morg}
M.~Morgenstern, C.~Wittneven, R.~Dombrowski, and R.~Wiesendanger, Phys.~Rev.~Lett. {\bf 84}, 5588 (2000).
\bibitem{Morgenstern}
M.~Morgenstern, J.~Klijn, C.~Meyer, and R.~Wiesendanger, Phys.~Rev.~Lett. {\bf 90}, 056804 (2003).
\bibitem{Hashimoto}
K.~Hashimoto, C.~Sohrmann, J.~Wiebe, T.~Inaoka, F.~Meier, Y.~Hirayama, R.~A.~R\"omer, R.~Wiesendanger, and M.~Morgenstern, Phys. Rev. Lett. {\bf 101}, 256802 (2008).
\bibitem{Hashimoto2}
K.~Hashimoto, T.~Champel, S.~Florens, C.~Sohrmann, J.~Wiebe, Y.~Hirayama, R.~A.~R\"omer, R.~Wiesendanger, and M.~Morgenstern, Phys. Rev. Lett. {\bf 109}, 116805 (2012).
\bibitem{Becker}
S.~Becker, C.~Karrasch, T.~Mashoff, M.~Pratzer, M.~Liebmann, V.~Meden, and M.~Morgenstern, Phys.~Rev.~Lett. {\bf 106}, 156805 (2011).
\bibitem{Andrei}
G.~Li, A.~Luican, and E.~Y.~Andrei, Phys.~Rev.~Lett. {\bf 102}, 176804 (2009).
\bibitem{Miller}
D.~L.~Miller, K.~D.~Kubista, G.~M.~Rutter, M.~Ruan, W.~A.~de Heer, P.~N.~First, and J.~A.~Stroscio, Science {\bf 324}, 924 (2009).
\bibitem{Song}
Y.~J.~Song, A.~F.~Otte, Y.~Kuk, Y.~Hu, D.~B.~Torrance, P.~N.~First, W.~A.~de Heer, H.~Min, S.~Adam, M.~D.~Stiles, A.~H.~MacDonald, and J.~A.~Stroscio, Nature {\bf 467}, 185 (2010).
\bibitem{Cheng}
P.~Cheng, C.~Song, T.~Zhang, Y.~Zhang, Y.~Wang, J-F.~Jia, J.~Wang, Y.~Wang, B-F.~Zhu, X.~Chen, X.~Ma, K.~He, L.~Wang, X.~Dai, Z.~Fang, X.~Xie, X-L.~Qi, C-X.~Liu, S-C.~Zhang, and Q-K.~Xue, Phys.~Rev.~Lett. {\bf 105}, 076801 (2010).
\bibitem{Hanaguri}
T.~Hanaguri, K.~Igarashi, M.~Kawamura, H.~Takagi, and T.~Sasagawa, Phys.~Rev.~B {\bf 82}, 081305(R) (2010).
\bibitem{Okada}
Y.~Okada, W.~Zhou, C.~Dhital, D.~Walkup, Y.~Ran, Z.~Wang, S.~D.~Wilson, and V.~Madhavan, Phys.~Rev.~Lett. {\bf 109}, 166407 (2012).
\bibitem{Jiang}
Y.~Jiang, Y.~Wang, M.~Chen, Z.~Li, C.~Song, K.~He, L.~Wang, X.~Chen, X.~Ma, and Q-K.~Xue, Phys.~Rev.~Lett. {\bf 108}, 016401 (2012).
\bibitem{Fu4}
Y-S.~Fu, M.~Kawamura, K.~Igarashi, H.~Takagi, T.~Hanaguri, and T.~Sasagawa, Nature Phys. {\bf 10}, 815 (2014).
\bibitem{Okada2}
Y.~Okada, M.~Serbyn, H.~Lin, D.~Walkup, W.~Zhou, C.~Dhital, M.~Neupane, S.~Xu, Y.~J.~Wang, R.~Sankar, F.~Chou, A.~Bansil, M.~Z.~Hasan, S.~D.~Wilson, L.~Fu, and V.~Madhavan, Science {\bf 341}, 1496 (2013).
\bibitem{Zeljikovic}
I. Zeljkovic, Y.~Okada, M.~Serbyn, R.~Sankar, D.~Walkup, W.~Zhou, J.~Liu, G.~Chang, Y.~J.~Wang, M.~Z.~Hasan, F.~Chou, H.~Lin, A.~Bansil, L.~Fu, and V.~Madhavan,  Nature Phys. {\bf 14}, 318 (2015).
\bibitem{Jeon}
S.~Jeon, B.~B.~Zhou, A.~Gyenis, B.~E.~Feldmann, I.~Kimichi, A.~C.~Potter, Q.~D.~Gibson, R.~J.~Cava, A.~Vishwanath, and A.~Yazdani, Nature Mat. {\bf 13}, 851 (2014).
\bibitem{Lencer}
D.~Lencer, M.~Salinga, B.~Grabowski, T.~Hickel, J.~Neugebauer, and M.~Wuttig, Nature Mat. {\bf 7}, 972 (2008).
\bibitem{Pauly1}
C.~Pauly, G.~Bihlmayer, M.~Liebmann, M.~Grob, A.~Georgi, D.~Subramaniam, M.~R.~Scholz, J.~Sanchez-Barriga, A.~Varykhalov, S.~Bl\"ugel, O.~Rader, and M.~Morgenstern, Phys.~Rev.~B {\bf 86}, 235106 (2012).
\bibitem{Plucinski}
L.~Plucinski, A.~Herdt, S.~Fahrendorf, G.~Bihlmayer, G.~Mussler, S.~D\"oring, J.~Kampmeier, F.~Matthes, D.~E.~B\"urgler, D.~Gr\"utzmacher, S.~Bl\"ugel, and C.~M.~Schneider, J.~Appl.~Phys. {\bf 113}, 053706 (2013).
\bibitem{Wang}
G.~Wang, X.~Zhu, J.~Wen, X.~Chen, K.~He, L.~Wang, X.~Ma, Y.~Liu, X.~Dai, Z.~Fang, J.~Jia, and Q.~Xue, Nano Res. {\bf 3}, 874 (2010).
\bibitem{Seibel}
C.~Seibel, H.~Maa{\ss}, M.~Ohtaka, S.~Fiedler, C.~J\"unger, C-H.~Min, H.~Bentmann, K.~Sakamoto, and F.~Reinert, Phys.~Rev.~B {\bf 86}, 161105(R) (2012).
\bibitem{Schwartz}
H.~Schwartz, G.~Bj\"orck, and O.~Beckman, Solid State Comm. {\bf 5}, 905 (1967).
\bibitem{Jiang1}
Y.~Jiang, Y.~Y.~Sun, M.~Chen, Y.~Wang, Z.~Li, C.~Song, K.~He, L.~Wang, X.~Chen, Q-K.~Xue, X.~Ma, and S.~B.~Zhang, Phys.~Rev.~Lett. {\bf 108}, 066809 (2012).
\bibitem{Hsieh2}
D.~Hsieh, Y.~Xia, D.~Qian, L.~Wray, F.~Meier, J.~H.~Dil, J.~Osterwalder, L.~Patthey, A.~V.~Fedorov, H.~Lin, A.~Bansil, D.~Grauer, Y.~S.~Hor, R.~J.~Cava, and M.~Z.~Hasan, Phys.~Rev.~Lett. {\bf 103}, 146401 (2009).
\bibitem{Mashoff}
T.~Mashoff, M.~Pratzer, and M.~Morgenstern, Rev.~Sci.~Instrum. {\bf 80}, 53702 (2009).
\bibitem{Morgenstern2}
M.~Morgenstern, Surf.~Rev.~Lett. {\bf 10}, 933 (2003).
\bibitem{Takagaki}
Y.~Takagaki, A.~Giussani, K.~Perumal, R.~Calarco, and K.-J.~Friedland, Phys.~Rev.~B {\bf 86}, 125137 (2012).
\bibitem{Peranio}
N.~Peranio, M.~Winkler, Z.~Aabdin, J.~K\"onig, H.~B\"ottner, and O.~Eibl, Phys.~Stat. Sol. A {\bf 209}, 289 (2012).
\bibitem{Castro}
A.~H.~Castro Neto, F.~Guinea, N.~M.~R.~Peres, K.~S.~Novoselov, and A.~K.~Geim, Rev.~Mod.~Phys. {\bf 81}, 109 (2009).
\bibitem{Seibel1}
C.~Seibel, H.~Bentmann, J.~Braun, J.~Minar, H.~Maa{\ss}, K.~Sakamoto, M.~Arita, K.~Shimada, H.~Ebert, and F.~Reinert, Phys.~Rev.~Lett. {\bf 114}, 066802 (2015).
\bibitem{Zhang3}
W.~Zhang, R.~Yu, H-J.~Zhang, X.~Dai, and Z.~Fang, New J.~Phys. {\bf 12}, 065013 (2010).
\bibitem{Shklovskii}
B. Skinner, T.~Chen, and B.~I.~Shklovskii, Phys. Rev. Lett. {\bf 109}, 176801 (2012).
\bibitem{Roth}
L.~M.~Roth, Phys.~Rev. {\bf 145}, 434 (1966).
\bibitem{Wright}
A.~R.~Wright and R.~H.~McKenzie, Phys.~Rev.~B {\bf 87}, 085411 (2013).
\bibitem{Xia}
Y.~Xia, D.~Qian, D.~Hsieh, L.~Wray, A.~Pal, H.~Lin, A.~Bansil, D.~Grauer, Y.~S.~Hor, R.~J.~Cava, and M.~Z.~Hasan, Nature Phys. {\bf 5}, 398 (2009).
\bibitem{Ando}
T. Ando, J. Phys. Soc. Jpn. {\bf 53}, 3101 (1984).
\bibitem{Champel}
T. Champel and S. Florens, Phys. Rev. B {\bf 75},
245326 (2007).
\bibitem{Grimvall}
G.~Grimvall, {\it The Electron-Phonon Interaction in Metals, Selected  Topics  in  Solid  State  Physics}, edited by E.~P.~Wohlfarth (North-Holland, New York, 1981).
\bibitem{Echenique}
P.~M.~Echenique, R.~Berndt, E.~V.~Chulkov, T.~Fauster, A.~Goldmann, and U.~H\"ofer, Surf. Sci. Rep. {\bf 52}, 219 (2004).
\bibitem{Sosso}
G.~C.~Sosso, S.~Caravati, and M.~Bernasconi, J. Phys.: Cond. Mat. {\bf 21}, 095410 (2009).
\bibitem{Richter}
W.~Richter, H.~K\"ohler, and C.~R.~Becker, Phys.~Stat. Sol. B {\bf 84}, 619 (1977).
\bibitem{Burgi}
L.~B\"urgi, O.~Jeandupeux, H.~Brune, and K.~Kern, Phys.~Rev.~Lett. {\bf 82}, 4516 (1999).
\bibitem{Gonzales}
J.~Gonzalez, F.~Guinea, and M.~A.~H.~Vozmediano, Phys. Rev. Lett. {\bf 77}, 3589 (1996).
\bibitem{Hwang}
E.~H.~Hwang, B.~Y-K.~Hu, and S.~Das Sarma, Phys. Rev. B {\bf 76}, 115434 (2007).
\bibitem{Sanchez2}
J.~Sanchez-Barriga, E.~Golias, A.~Varykhalov, J.~Braun, L.~V.~Yashina, R.~Schumann, J.~Minar, H.~Ebert, O.~Kornilov, and O.~Rader, arXiv:1505.02742 (2015).
\bibitem{Reimann}
J.~Reimann, J.~G\"udde, K.~Kuroda, E.~V.~Chulkov, and U.~H\"ofer, Phys. Rev. B {\bf 90}, 081106 (2014).
\bibitem{Sanchez}
J. Sanchez-Barriga, M.~R.~Scholz, E.~Golias, E.~Rienks, D.~Marchenko, A.~Varykhalov, L.~V.~Yashina, and O.~Rader, Phys. Rev. B {\bf 90}, 195413 (2014).
\bibitem{Chen}
C.~Chen, Z.~Xie, Y.~Feng, H.~Yi, A.~Liang, S.~He, D.~Mou, J.~He, Y.~Peng, X.~Liu, Y.~Liu, L.~Zhao, G.~Liu, X.~Dong, J.~Zhang, L.~Yu, X.~Wang, Q.~Peng, Z.~Wang, S.~Zhang, F.~Yang, C.~Chen, Z.~Xu, and X.~J.~Zhou, Sci. Rep. 3, {\bf 2411} (2013).
\bibitem{Hajlaoui}
M.~Hajlaoui, E.~Papalazarou, J.~Mauchain, G.~Lantz, N.~Moisan, D.~Boschetto, Z.~Jiang, I.~Miotkowski, Y.~P.~Chen, A.~Taleb-Ibrahimi, L.~Perfetti, and M.~Marsi, Nano Lett. {\bf 12}, 3532 (2012).
\bibitem{Crepaldi}
A. Crepaldi, B.~Ressel, F.~Cilento, M.~Zacchigna, C.~Grazioli, H.~Berger, P.~Bugnon, K.~Kern, M.~Grioni, and F.~Parmigiani, Phys. Rev. B {\bf 86}, 205133 (2012).
\bibitem{Song2}
C-L.~Song, L.~Wang, K.~He, S-H.~Ji, X.~Chen, X-C.~Ma, and Q-K.~Xue, arXiv:1504.03397v1.
\end{thebibliography}
\end{document}